\documentclass[onecolumn,conference]{IEEEtran}

\usepackage{cite}
\usepackage{amsmath,amssymb}
\usepackage{graphicx}
\usepackage{xcolor}
\usepackage{booktabs}
\usepackage{multirow}
\usepackage{hyperref}

\begin{document}


\title{A Reconfigurable Hybrid Convolutional–Fully Connected Neuromorphic Core for Biomedical Edge Inference}

\author{\IEEEauthorblockN{Sarah Johari, Suman Kumar, Abhishek Mishra, Anush Lingamoorthy, and Nagarajan Kandasamy}
\IEEEauthorblockA{Electrical and Computer Engineering Department \\
Drexel University \\
Philadelphia, Pennsylvania 19104, USA \\
\{sj984,sk4266,am4862,kandasamy\}@drexel.edu
}}


\maketitle

\begin{abstract}
This work presents a programmable FPGA-based architecture for spiking convolutional neural network (SCNN) inference, with real-time hypoxia classification serving as a biomedical edge application. The architecture implements a hybrid spiking convolutional--fully connected (CNN--FC) topology on a programmable, quantized, layer-based neuromorphic hardware core. Early layers perform spiking convolution using receptive-field connectivity with support for multi-channel kernels and stride, while deeper layers use fully connected spiking layers for classification. A PyTorch-based hardware--software co-design flow enables deployment of trained parameters with quantization and configurability support. The design is first validated on MNIST and Fashion-MNIST, achieving hardware accuracies of up to 98\% and 86\%, respectively, at 16-bit precision. It is then applied to hypoxia classification using red and infrared photoplethysmography (PPG) signals acquired from a shoulder-mounted sensor, with skin tone included as an additional input channel. The resulting classifier achieves an average hardware accuracy of 88.26\% across five folds at 16-bit precision while consuming 1.455~W of dynamic power, demonstrating the feasibility of low-power neuromorphic biomedical classification at the edge.
\end{abstract}


\section{Introduction}\label{sec:introduction}


Spiking neural networks (SNNs) process information using discrete spikes, which can substantially reduce power consumption when executed on dedicated neuromorphic hardware. Recently, FPGA-based neuromorphic designs have been proposed, from small-scale fully connected networks to architectures aimed at real-world inference~\cite{matinizadeh2024towards,10942719,gautam2025neurocorex,mohammadhassani2025improving,10658962}. 

SNNs on neuromorphic hardware are an attractive solution for biomedical applications, where continuous physiological monitoring requires low-power real-time inference. A representative and important example is the detection of hypoxia, through the continuous estimation of blood oxygen saturation (SpO\textsubscript{2}), which is central to a wide range of clinical and community health settings. In acute care, SpO\textsubscript{2} monitoring guides oxygen therapy in intensive care, post-surgical recovery, and mechanical ventilation. In sleep medicine, it is the standard method for detecting obstructive sleep apnea, where episodic desaturation is associated with cardiovascular risk. In high-altitude medicine, it helps identify people at risk for hypoxemia, pulmonary edema, and cerebral edema. Finally, opioid overdose detection is one of the most time-critical applications: nearly 40\% of overdose cases progress to multiorgan failure due to prolonged oxygen deprivation, and the window to prevent irreversible brain injury is often under five minutes~\cite{roth2025}.

Recent work has demonstrated the effectiveness of SNNs in classifying the severity of hypoxia through a non-invasive method using photoplethysmography (PPG) signals~\cite{lingamoorthy2026hypoxspike}. However, these models have yet to be deployed on dedicated neuromorphic hardware, leaving a gap between algorithmic capability and practical edge deployment.

Returning to the available neuromorphic hardware designs, most focus on fully connected (FC) topologies, leaving convolutional SNN accelerators---critical for feature extraction in image and signal classification---underexplored, particularly with respect to configurability and quantization support~\cite{carpegna2022spiker,neil2014minitaur}. Systolic array architectures offer a promising avenue for convolutional SNN acceleration. For example, Guo et al.~\cite{guo2019systolic} showed that binary spike activations are well-suited to systolic arrays, where multiply-accumulate operations reduce to additions. Lee and Li~\cite{lee2020reconfigurable} demonstrated the impact of variable tiling on throughput, Lee et al.~\cite{lee2022saarsp} introduced temporal parallelism for recurrent SNNs, and Wang et al.~\cite{wang2025spikeflow} proposed SpikeFlow with adjustable temporal-spatial dataflow. However, these implementations have largely focused on fixed topologies with limited reconfigurability, and their deployment with full hardware-software co-design and quantization support remains underexplored for hybrid CNN--FC frameworks targeting biomedical edge applications.

Building on convolutional feature extraction, Aung et al.\ proposed DeepFire~\cite{aung2021deepfire} and DeepFire2~\cite{aung2023deepfire2} as dedicated accelerators for spiking CNN inference, but neither addresses hybrid CNN--FC topologies with layer-wise reconfigurability, mixed-precision quantization. 

We develop a neuromorphic hardware platform for end-to-end spiking CNN inference for hypoxia detection. The main contributions of this work are:
\begin{itemize}
    \item A programmable, quantized digital spiking CNN accelerator for \emph{biomedical edge application} that supports hybrid CNN--FC topologies with multi-level configurability, including arithmetic precision, kernel size, stride, neuron parameters, and layer composition.

    \item We formulate hypoxia detection as a classification task (based on the severity of the condition) using red and infrared PPG signals along with skin tone as input channels. We then develop and deploy a spiking CNN on the proposed hardware accelerator.
    
    \item To validate the generalizability of the design, we tested it on image-based benchmarks (MNIST and Fashion-MNIST) to demonstrate competitive classification accuracy under both 8-bit and 16-bit quantization levels.
\end{itemize}

The remainder of this paper is organized as follows. Section~\ref{sec:background} provides the necessary background. Section~\ref{sec:architecture} develops the hybrid CNN--FC spiking hardware and Section~\ref{sec:hypoxia} discusses the deployment of an SNN trained to detect hypoxia on this hardware. Section~\ref{sec:conclusion} concludes the paper.

\section{Background}\label{sec:background}

This section presents background on spiking neural networks (SNNs) and surveys open-source neuromorphic platforms that support their execution. It also discusses the importance of detecting hypoxia at the edge, which motivates the deployment of hypoxia detection and classification algorithms on neuromorphic hardware.

\subsection{Spiking Neurons and Network Models}
SNNs are biologically plausible computing models in which neurons communicate through discrete spikes rather than continuous-valued activations. The Leaky Integrate-and-Fire (LIF) neuron has become the predominant choice for digital hardware implementations due to its favorable trade-off between biological fidelity and computational simplicity~\cite{10405978,`10242251,10766711,11043873}. The continuous-time dynamics of the neuron neuron is governed by the first-order ordinary differential equation
\begin{equation}
\tau \frac{dV_{\text{mem}}(t)}{dt} = -V_{\text{mem}}(t) + R \cdot I_{\text{in}}(t),
\label{eq:lif_ode}
\end{equation}
where $V_{\text{mem}}(t)$ is the membrane potential, $\tau = R \cdot C$ is the membrane time constant defined by the membrane resistance $R$ and capacitance $C$, and $I_{\text{in}}(t)$ is the input current~\cite{10658962}. Applying the forward Euler discretization yields
\begin{equation}
V_{\text{mem}}[t+1] = V_{\text{mem}}[t] - \underbrace{\frac{\Delta t}{\tau}}_{\text{decay\_rate}} \cdot V_{\text{mem}}[t] + \underbrace{\frac{\Delta t}{C}}_{\text{growth\_rate}} \cdot I_{\text{in}}[t].
\label{eq:lif_discrete}
\end{equation}
When $V_{\text{mem}}[t]$ exceeds a threshold voltage $V_{\text{th}}$, the neuron emits a spike and the membrane potential is reset according to a configurable reset mechanism (e.g., reset-to-zero, reset-by-subtraction, or reset-to-constant). In addition, a refractory period may be enforced to limit the maximum firing frequency.

For current-based synapses, the input current to a post-synaptic neuron $j$ is calculated as the weighted sum of incoming spikes  $I_j[t] = \sum_{i} x_{ij}[t] \cdot w_{ij}$,
where $x_{ij}[t] \in \{0, 1\}$ is the binary spike from pre-synaptic neuron $i$ and $w_{ij}$ is the corresponding synaptic weight. The binary nature of spikes reduces the multiply-accumulate operation to a conditional addition, which is highly advantageous for digital hardware implementations.

\begin{figure}[t!]
    \centering
    \includegraphics[width=0.72\textwidth]{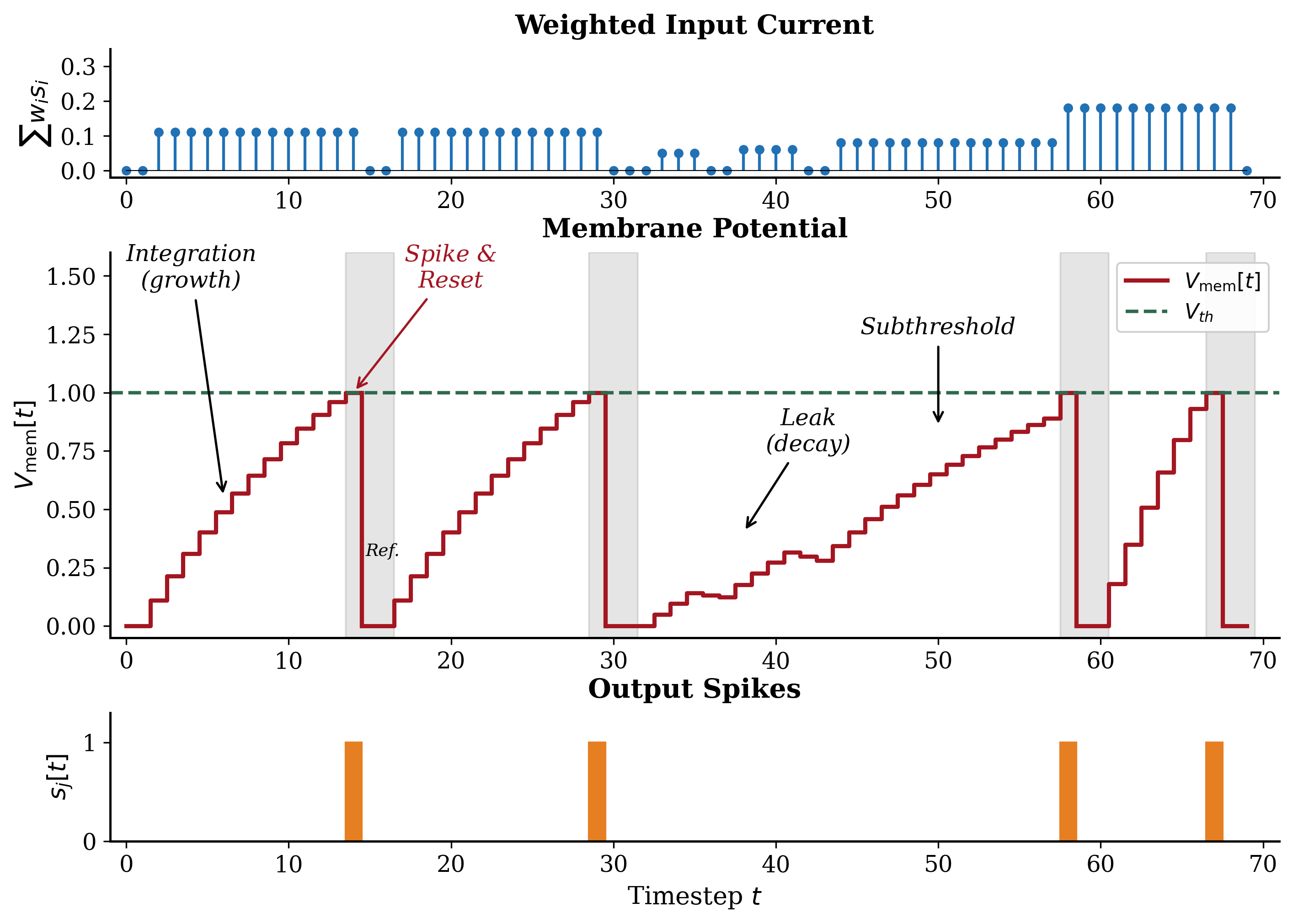}
    \caption{Dynamics of a digital LIF neuron showing integration, threshold crossing, spike-and-reset, leak decay, and subthreshold activity.}
    \label{fig:lif_dynamics}
    \vspace{-12pt}
\end{figure}

When multiple LIF neurons are organized into layers with all-to-all connectivity, the result is a fully connected spiking network suitable for classification tasks~\cite{matinizadeh2024fully,10766554}. However, FC topologies do not scale efficiently to high-dimensional inputs, as the number of synaptic connections grows quadratically with the input size. To address this, convolutional architectures have been adapted to the spiking domain~\cite{cao2015spiking,aung2021deepfire}, where shared kernels are applied across spike-encoded input feature maps and the weighted sums of binary spike inputs are accumulated into the membrane potentials of post-synaptic LIF neurons. This preserves the parameter efficiency of conventional CNNs while benefiting from the sparse, event-driven computation of SNNs. A typical hybrid spiking CNN--FC pipeline consists of one or more convolutional layers for feature extraction, followed by fully connected layers for classification, with the predicted class corresponding to the output neuron with the highest accumulated spike count.

\subsection{Open-Source Availability and Research Gap}
Deploying SNNs on hardware requires an accessible and reproducible platform. Although SNN accelerators based on FPGA and ASIC have received increasing attention, only a small number of such platforms are publicly available as open source~\cite{isik2023survey,carpegna2022spiker,modaresi2023openspike}. The majority of reported implementations remain closed source, limiting reproducibility and preventing fair comparison across designs. In addition, no existing open-source platform supports a fully configurable hybrid CNN--FC spiking architecture with variable quantization and a complete hardware--software co-design flow, which motivates the present work.

\subsection{Hypoxia Detection using Machine Learning}
Hypoxia is a physiological state in which the body, or a specific tissue region, is deprived of adequate oxygen. Clinically, it is commonly associated with peripheral oxygen saturation (SpO\textsubscript{2}) levels below 92\% in healthy individuals and below 88\% in patients with chronic obstructive pulmonary disease~\cite{sjoding2020racial}. If not treated, sustained oxygen deprivation leads to progressive metabolic failure, neurological injury, and possible death. Consequently, timely detection of hypoxia is critical in a wide range of clinical and community health settings.

We focus on SpO\textsubscript{2} as the physiological signal used to detect hypoxia. SpO\textsubscript{2} reflects the oxygen-carrying state of hemoglobin in peripheral capillary blood and can be estimated noninvasively using photoplethysmography (PPG), an optical sensing technique in which light at specific wavelengths is emitted into the skin and the reflected or transmitted signal is measured by a photodetector. The estimation relies on the differential absorption of red and infrared light as it passes through tissue. However, these measurements can also be influenced by skin pigmentation, which may reduce accuracy across diverse patient populations~\cite{sjoding2020racial}. To address this limitation, recent studies have explored machine learning approaches that incorporate skin tone as an additional input feature alongside raw PPG waveforms, including approaches based on spiking neural networks. For example, Lingamoorthy et al.~\cite{lingamoorthy2026hypoxspike} proposed HypoxSpike, an SNN that classifies three levels of hypoxemia (normal, moderate or severe) using PPG signals, motion data, and skin tone collected from a shoulder-mounted sensor. However, HypoxSpike was evaluated only in software, with hardware inference left as future work.


\section{Hardware Design}\label{sec:architecture}
\label{sec:architecture}
This section develops the hardware architecture of the proposed spiking convolutional neural network (SCNN) core. The design builds upon the QUANTISENC neuromorphic platform ~\cite{matinizadeh2024fully}, one of the few design available as open source, and extends it with convolutional layer support, yielding a unified hybrid CNN--FC architecture. The section first describes the foundational building blocks, the quantized arithmetic units and the LIF neuron, then details the fully connected and convolutional layer implementations, and finally presents the top-level system integration.

\subsection{LIF Core}
Arithmetic within the SCNN core uses signed fixed-point representation with configurable precision $Q_{n.q}$, where $n$ and $q$ denote the number of integer and fractional bits, respectively. Including the sign bit, neuron state variables therefore use a total word width of $1+n+q$ bits. Synaptic weights can be represented using $1+m+q$ bits, with $m \leq n$, to reduce the memory footprint while retaining an adequate dynamic range for trained values. The core computation is built on three arithmetic modules: (i) a quantized adder with overflow/underflow saturation, (ii) a quantized multiplier that forms a $2N$-bit intermediate result and requantizes it to $N$ bits, and (iii) a quantized ALU that additionally provides comparator support for threshold detection~\cite{matinizadeh2024fully}.

\begin{figure}[t!]
    \centering
    \includegraphics[width=0.62\textwidth]{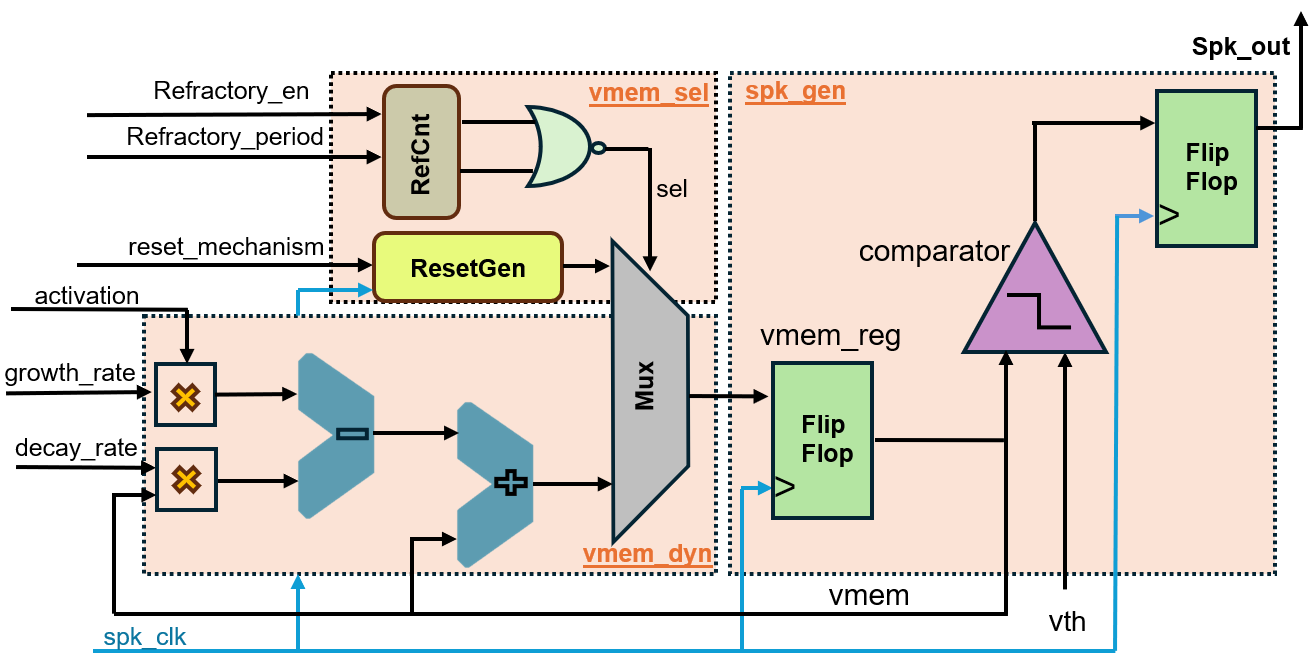}
    \caption{Hardware block diagram of the LIF neuron.}
    \label{fig:LIF}
    \vspace{-12pt}
\end{figure}

The LIF neuron is the core computational unit of the architecture. Each neuron maintains an internal membrane potential register, \texttt{vmem}, and implements the discrete-time dynamics of Eq.~\eqref{eq:lif_discrete}, including configurable decay and growth, threshold comparison, three reset modes, and a configurable refractory period. All neuron parameters are stored in runtime-configurable registers, allowing post-deployment tuning without resynthesis. As shown in Fig.~\ref{fig:LIF}, the LIF block operates in the spike-clock domain (\texttt{spk\_clk}). The incoming activation (\texttt{activation}) from synaptic memory is multiplied by \texttt{growth\_rate}, while the current membrane potential (\texttt{vmem}) is multiplied by \texttt{decay\_rate}, forming two parallel datapaths whose outputs are combined through quantized adders to generate the candidate membrane update (\texttt{vmem\_dyn}). At the same time, the ResetGen block computes the post-spike membrane value based on the configured \texttt{reset\_mechanism}, and the refractory counter (RefCnt) tracks whether the neuron is in its refractory phase. A multiplexer, controlled by \texttt{vmem\_sel} derived from the RefCnt and comparator outputs, selects between the normal update path and the reset path. The selected value is stored in the \texttt{vmem\_reg} flip-flop, completing the update. Finally, the comparator checks \texttt{vmem} against \texttt{vth} and drives the spike-generation logic (\texttt{spk\_gen}), which produces the output spike (\texttt{Spk\_out}) through a final registered stage~\cite{matinizadeh2024fully}.

\subsection{Synaptic Memory and Multiply-Accumulate}
Synaptic memory and multiply-accumulate (MAC) operations are organized differently for fully connected and convolutional layers, reflecting their distinct connectivity patterns.

\begin{figure}[t!]
    \centering
    \includegraphics[width=0.90\textwidth]{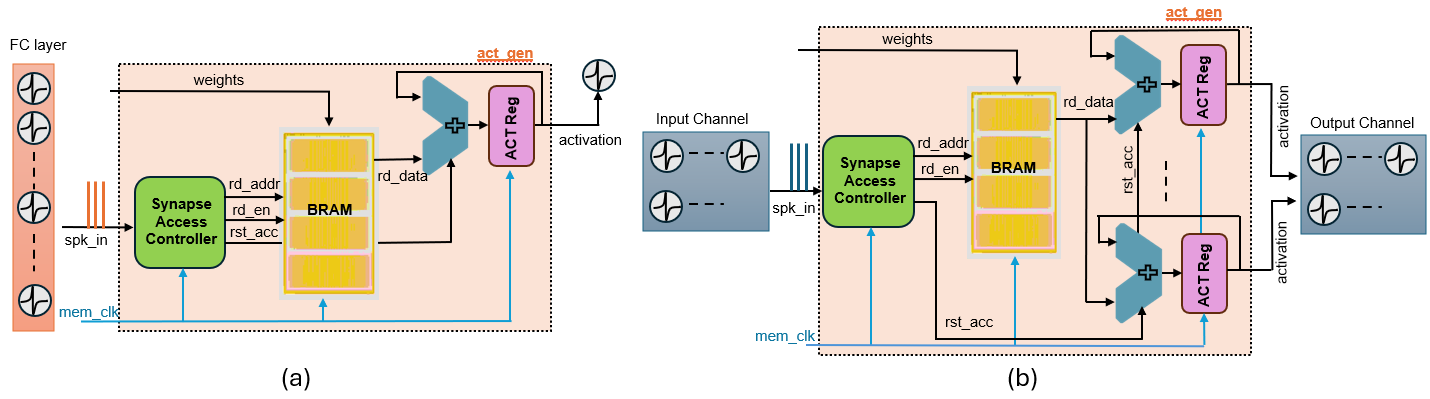}
    \caption{Hardware block diagram of multiply-accumulate (MAC) operations:
    (a) Fully connected layer and (b) convolutional (CNN) layer.}
    \label{fig:MAC_fc_cnn}
\end{figure}

\begin{figure}[t!]
    \centering
    \includegraphics[width=0.85\textwidth]{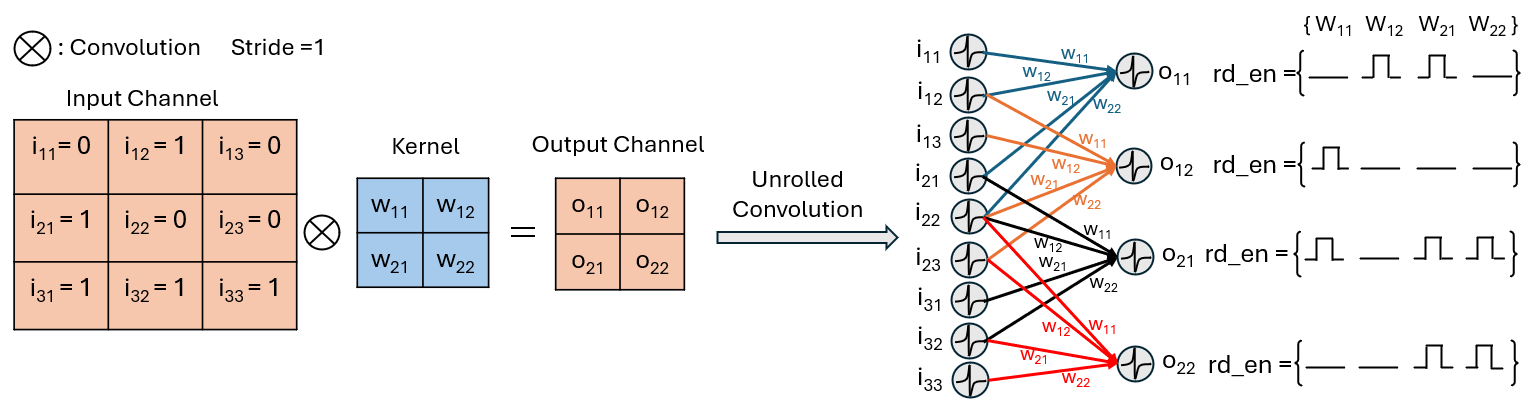}
    \caption{Unrolled convolution example: a $3 \times 3$ input convolved with a $2 \times 2$ shared kernel (stride = 1) producing a $2 \times 2$ output. Each output neuron receives connections from its receptive field, with the shared kernel counter generating per-output \texttt{rd\_en} signals gated by the input spike pattern.}
    \label{fig:unrolled_conv}
    \vspace{-12pt}
\end{figure}

\subsubsection{Fully Connected Synaptic Access}
In the FC layer, each post-synaptic neuron maintains a dedicated block RAM storing $F$ synaptic weights, where $F$ is the fan-in. As shown in Fig.~\ref{fig:MAC_fc_cnn}(a), a synapse access controller generates the read address (\texttt{rd\_addr}), read enable (\texttt{rd\_en}), and accumulator reset (\texttt{rst\_acc}) signals based on the input spike pattern (\texttt{spk\_in}). The controller sequentially scans through the pre-synaptic addresses, asserting a read enable only when the corresponding input spike is active, thereby implementing the current accumulation. Each weight read from BRAM (\texttt{rd\_data}) is conditionally added to a running partial sum in the activation register (ACT~Reg), producing the total synaptic activation forwarded to the LIF block~\cite{matinizadeh2024fully}.

\subsubsection{Convolutional Synaptic Access}
The convolutional layer adopts a memory organization based on kernel weight sharing. Rather than storing a full weight matrix per neuron, each input--output channel pair shares a single block RAM of size $K_x \times K_y$. All output feature-map positions for that channel pair access the same kernel memory, greatly reducing memory cost compared to an equivalent fully connected design.

The convolutional synaptic access controller implements an unrolled convolution in which all output positions are computed in parallel. A shared counter $k$ steps through the kernel entries from $0$ to ($K_x \times K_y - 1$), reading one weight per cycle from the shared BRAM. In each cycle, $k$ is decomposed into row and column offsets, $k_x$ and $k_y$. For each output position $(w_x, w_y)$, the controller then maps this kernel offset to the corresponding input neuron as 
\begin{equation}
\text{idx} = (w_x \cdot S + k_x) \cdot Y_{\text{in}} + (w_y \cdot S + k_y),
\label{eq:receptive_field}
\end{equation}
where $S$ is the stride and $Y_{\text{in}}$ is the input width. If the input spike at this index is active, the read enable for that output position is asserted, and the shared kernel weight is added to that position's partial sum accumulator. Since all $X_{\text{out}} \times Y_{\text{out}}$ output positions evaluate their respective input indices in the same clock cycle, the entire convolution completes in just $K_x \times K_y$ memory clock cycles---one cycle per kernel entry.

Figure~\ref{fig:unrolled_conv} illustrates the process with a concrete example. A $3 \times 3$ input feature map is convolved with a $2 \times 2$ kernel with stride  = 1, yielding a $2 \times 2$ output. The left side shows the standard convolution, and the right side shows its unrolled hardware realization. Each output neuron ($o_{11}$, $o_{12}$, $o_{21}$, $o_{22}$) is connected to its corresponding $2 \times 2$ receptive field, while all four outputs share the same kernel weights, ${w_{11}, w_{12}, w_{21}, w_{22}}$, stored in a single BRAM. A shared counter traverses the four kernel entries over four clock cycles. At each step, \texttt{rd\_en} is asserted for an output only when the corresponding input neuron spikes. For example, at $k=0$ (weight $w_{11}$), output $o_{11}$ reads input $i_{11}$, which is 0, so no accumulation occurs, while output $o_{12}$ reads input $i_{12}$, which is 1, so \texttt{rd\_en} is asserted and $w_{11}$ is accumulated. The \texttt{rd\_en} waveforms beside each output neuron visualize this spike-gated accumulation across the kernel sweep.


As shown in Fig.~\ref{fig:MAC_fc_cnn}(b), each output position maintains its own independent partial sum register. When a new input spike pattern arrives, all accumulators are reset and a new kernel sweep begins. After the sweep completes, the accumulated partial sums represent the full convolution result and are forwarded to the LIF neurons for membrane potential integration.

\subsection{Top-Level System Integration}

The SCNN core integrates multiple CNN layers followed by FC classification layers into a unified processing pipeline. The layer configuration is specified via a compile-time parameter array that defines the kernel dimensions, stride, and number of input/output channels for each convolutional layer. Output spikes from last CNN layer are flattened and forwarded as the input spike vector to the fully-connected layer.

A layer-level address decoder selects the target layer for weight programming based on a dedicated field in the write address bus. All layers share a common set of neuron configuration registers (threshold, decay rate, growth rate, resting potential, reset mechanism, and refractory period), which are programmed at runtime through the configuration interface. The architecture operates on a dual-clock scheme: a fast memory clock (\texttt{memclk}) drives the synaptic weight access and MAC operations, while a slower spike clock (\texttt{spkclk}) governs the LIF neuron dynamics, with clock domain crossing handled by the synchronizer modules.

\subsection{Design Space and Configurability}
The proposed architecture offers configurability at multiple levels, summarized in Table~\ref{tab:config_params}. At the \emph{arithmetic level}, the integer and fractional bit widths can be set independently, enabling mixed-precision deployment. At the \emph{layer level}, kernel dimensions, stride, and the number of input/output channels are parameterized, supporting a wide range of convolutional topologies. At the \emph{neuron level}, all LIF parameters are configurable via memory-mapped registers. At the \emph{system level}, the number of CNN and FC layers, as well as their interconnection, are defined at compile time through the configuration package. This multi-level configurability enables rapid design space exploration and efficient deployment across diverse application scenarios.

\begin{table}[t!]
    \centering
    \caption{Summary of configurable parameters across different levels of the proposed architecture.}
    \label{tab:config_params}
    \begin{tabular}{@{}llp{3.2cm}@{}}
        \toprule
        \textbf{Level} & \textbf{Parameter} & \textbf{Description} \\
        \midrule
        Arithmetic & \texttt{INTEGER\_PRECISION}      & Number of states integer bits ($n$) \\
                   & \texttt{DECIMAL\_PRECISION}      & Number of fraction bits ($q$) \\
                   & \texttt{WT\_INETGER\_PRECISION}  & Number of weights integer bits ($m$) \\
        \midrule
        Neuron & \texttt{VTH}                & Threshold voltage \\
               & \texttt{DECAY\_RATE}        & Membrane decay rate \\
               & \texttt{GROWTH\_RATE}       & Membrane growth rate \\
               & \texttt{VREST}              & Resting potential \\
               & \texttt{RESET\_MECHANISM}   & Reset mode (0, 1, or 2) \\
               & \texttt{REFRECTORY\_PERIOD} & Refractory duration \\
        \midrule
        FC Layer & \texttt{FANIN}  & Pre-synaptic fan-in \\
                 & \texttt{FANOUT} & Number of output neurons \\
        \midrule
        CNN Layer & \texttt{X\_KERNEL}, \texttt{Y\_KERNEL} & Kernel dimensions \\
                  & \texttt{STRIDE}                        & Convolution stride \\
                  & \texttt{IN\_CHANNEL}                   & Number of input channels \\
                  & \texttt{OUT\_CHANNEL}                  & Number of output channels \\
                  & \texttt{X\_FANIN}, \texttt{Y\_FANIN}  & Input feature map size \\
        \midrule
        System & \texttt{NUM\_CNN\_LAYERS}  & Number of CNN layers \\
               & \texttt{NUM\_FC\_LAYERS}   & Number of FC layers \\
               & \texttt{HARDWARE\_LAYERS}  & Total number of layers \\
        \bottomrule
    \end{tabular}
    \vspace{-12pt}
\end{table}

\subsection{Performance Evaluation using MNIST and FMNIST}
The hybrid CNN--FC spiking architecture was synthesized and evaluated on a Xilinx Virtex UltraScale+ FPGA. The networks were trained using snnTorch with a PyTorch backend, and the learned parameters were quantized and deployed onto the hardware via the co-design interface with per-layer quantization. Multiple fixed-point precision configurations were evaluated: 8-bit and 16-bit for MNIST, and 12-bit and 16-bit for FMNIST. Table~\ref{tab:img_arch} shows the network configurations.

\begin{table}[t!]
    \centering
    \caption{Spiking CNN--FC network configurations for MNIST and FMNIST.}
    \label{tab:img_arch}
    \begin{tabular}{llc}
        \toprule
        \textbf{Layer} & \textbf{Configuration} & \textbf{Output} \\
        \midrule
        \multicolumn{3}{l}{\textbf{MNIST}} \\
        \midrule
        Input          & 1 channel, $16 \times 16$     & 1$\times$16$\times$16 \\
        Conv 1 + LIF   & 16 out ch, kernel=5, stride=2 & 16$\times$6$\times$6 \\
        Conv 2 + LIF   & 32 out ch, kernel=5, stride=1 & 32$\times$2$\times$2 \\
        Flatten        & ---                           & 128 \\
        FC + LIF       & 10 units                      & 10 \\
        \midrule
        \multicolumn{3}{l}{\textbf{FMNIST}} \\
        \midrule
        Input          & 1 channel, $16 \times 16$     & 1$\times$16$\times$16 \\
        Conv 1 + LIF   & 16 out ch, kernel=5, stride=2 & 16$\times$6$\times$6 \\
        Conv 2 + LIF   & 32 out ch, kernel=3, stride=2 & 32$\times$2$\times$2 \\
        Flatten        & ---                           & 128 \\
        FC 1 + LIF     & 64 units                      & 64 \\
        FC 2 + LIF     & 10 units                      & 10 \\
        \bottomrule
    \end{tabular}
    \vspace{-6pt}
\end{table}

\begin{table}[t!]
    \centering
    \caption{Classification accuracy (\%) on MNIST and FMNIST.}
    \label{tab:img_accuracy}
    \begin{tabular}{llcc}
        \toprule
        \textbf{Task} & \textbf{Precision} & \textbf{SW} & \textbf{HW} \\
        \midrule
        \multirow{2}{*}{MNIST}
        & $Q_{0.4}/Q_{3.4}$ (8-bit)  & 98.5 & 97.0 \\
        & $Q_{0.8}/Q_{7.8}$ (16-bit) & 98.5 & 98.0 \\
        \midrule
        \multirow{2}{*}{FMNIST}
        & $Q_{1.6}/Q_{5.6}$ (12-bit)  & 86.5 & 81.5 \\
        & $Q_{1.8}/Q_{7.8}$ (16-bit) & 86.5 & 86.0 \\
        \bottomrule
    \end{tabular}
    \vspace{-6pt}
\end{table}

\begin{table}[t!]
    \centering
    \caption{Post-implementation FPGA resource utilization, dynamic power, and energy efficiency for MNIST and FMNIST at 0.5MHz.}
    \label{tab:img_util_power}
    \begin{tabular}{llrrrr}
        \toprule
        \textbf{Task} & \textbf{Metric} & \multicolumn{2}{c}{\textbf{Higher prec.}} & \multicolumn{2}{c}{\textbf{Lower prec.}} \\
        \midrule
        \multirow{7}{*}{MNIST}
        & & \multicolumn{2}{c}{\textbf{16-bit}} & \multicolumn{2}{c}{\textbf{8-bit}} \\
        \cmidrule(lr){3-4} \cmidrule(lr){5-6}
        & & \textbf{Used} & \textbf{\%} & \textbf{Used} & \textbf{\%} \\
        \cmidrule(lr){2-6}
        & LUT           & 357{,}717  & 27.44 & 269{,}190 & 20.65 \\
        & FF            & 194{,}067  & 7.44  & 109{,}504 & 4.20  \\
        & BRAM          & 269        & 13.34 & 269       & 13.34 \\
        & DSP           & 1{,}428    & 15.82 & ---       & ---   \\
        \cmidrule(lr){2-6}
        & Dynamic (W)   & \multicolumn{2}{c}{1.556} & \multicolumn{2}{c}{0.930} \\
        & MOPS/W        & \multicolumn{2}{c}{21.72} & \multicolumn{2}{c}{36.34} \\
        \midrule
        \multirow{7}{*}{FMNIST}
        & & \multicolumn{2}{c}{\textbf{16-bit}} & \multicolumn{2}{c}{\textbf{12-bit}} \\
        \cmidrule(lr){3-4} \cmidrule(lr){5-6}
        & & \textbf{Used} & \textbf{\%} & \textbf{Used} & \textbf{\%} \\
        \cmidrule(lr){2-6}
        & LUT           & 379{,}122  & 29.08 & 262{,}206 & 20.11 \\
        & FF            & 199{,}562  & 7.65  & 155{,}812 & 5.98  \\
        & BRAM          & 301        & 14.93 & 301       & 14.93 \\
        & DSP           & 1{,}556    & 17.24 & 1{,}556   & 17.24 \\
        \cmidrule(lr){2-6}
        & Dynamic (W)   & \multicolumn{2}{c}{1.746} & \multicolumn{2}{c}{1.421} \\
        & MOPS/W        & \multicolumn{2}{c}{12.13}  & \multicolumn{2}{c}{14.91} \\
        \bottomrule
    \end{tabular}
    \vspace{-12pt}
\end{table}

Table~\ref{tab:img_accuracy} summarizes the classification accuracies for MNIST and FMNIST. On MNIST, the 8-bit configuration achieves a hardware accuracy of 97.0\%, corresponding to a 1.5\% loss relative to the software baseline, while the 16-bit configuration reaches 98.0\%. On FMNIST, the 16-bit configuration attains 86.0\%, closely matching the 86.5\% software baseline, whereas the 12-bit configuration drops to 81.5\%. This indicates that FMNIST is a more challenging dataset and benefits more from higher numerical precision than MNIST.

Table~\ref{tab:img_util_power} reports the post-implementation FPGA resource utilization. For MNIST at 8-bit precision, the design occupies 20.65\% of LUTs and 4.20\% of FFs, increasing to 27.44\% and 7.44\% at 16-bit. FMNIST follows a similar trend with slightly higher utilization as a result of the additional FC layer. DSP slices are utilized in the 16-bit and 12-bit configurations for dedicated multiplication, while the 8-bit designs implement all arithmetic using LUT-based logic. BRAM usage remains constant across precisions for each task, as it is determined by the network topology rather than the quantization setting. For MNIST, the 8-bit configuration consumes 0.930~W of dynamic power and achieves 36.34~MOPS/W, while the 16-bit configuration requires 1.556~W at 21.72~MOPS/W --- a 1.67$\times$ improvement in energy efficiency at reduced precision. FMNIST shows a similar trend, with 12-bit achieving 14.91~MOPS/W at 1.421~W compared to 12.13~MOPS/W at 1.746~W for 16-bit.

\begin{figure}[t!]
    \centering
    \includegraphics[width=0.85\textwidth]{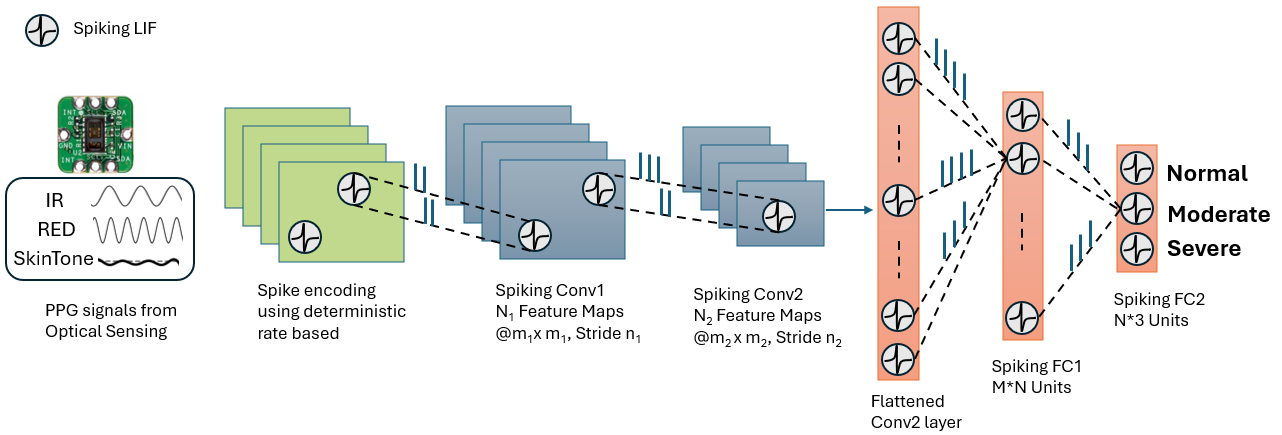}
    \caption{Architecture of the hybrid spiking CNN--FC network for classification of SpO\textsubscript{2} levels from PPG signals.}
    \label{fig:scnn_architecture}
    \vspace{-12pt}
\end{figure}

\section{Hypoxia Detection}\label{sec:hypoxia}
We formulate the problem as a three-class classification task using red PPG, infrared PPG, and skin tone as input channels to a spiking convolutional neural network, whose architecture is shown in Fig.~\ref{fig:scnn_architecture}. The classes are labeled according to the SpO\textsubscript{2} level: normal ($>$92\%), moderate (80--92\%) or severe ($<$80\%). The input signals are encoded into spike trains using  rate coding and processed through the proposed hybrid CNN--FC architecture, bridging the gap between software-based SNN classifiers and on-device neuromorphic deployment. The network configuration is summarized in Table~\ref{tab:spo2_arch}.

\begin{table}[t!]
    \centering
    \caption{Spiking CNN--FC network configuration for SpO2 classification.}
    \label{tab:spo2_arch}
    \begin{tabular}{llc}
        \toprule
        \textbf{Layer} & \textbf{Configuration} & \textbf{Output} \\
        \midrule
        Input          & 3 channels (Red, IR, Skin Tone) & 3 $\times$ 100 \\
        Conv 1 + LIF   & 16 out ch, kernel=5, stride=2 & 16 $\times$ 48 \\
        Conv 2 + LIF   & 32 out ch, kernel=5, stride=5 & 32 $\times$ 9 \\
        Flatten        & ---                           & 288 \\
        FC 1 + LIF     & 64 units                      & 64 \\
        FC 2 + LIF     & 3 units                       & 3 \\
        \midrule
        Output         & \multicolumn{2}{c}{Normal / Moderate / Severe} \\
        \bottomrule
    \end{tabular}
    \vspace{-12pt}
\end{table}

\subsection{Dataset Collection}
The dataset was collected at the UCSF Hypoxia Laboratory, a specialized research facility that induces precise, graded levels of hypoxemia under medical supervision. The study enrolled $N=12$ subjects and used a controlled protocol in which each participant’s SpO\textsubscript{2} was systematically varied from normoxic levels (~100\%) to severe hypoxemia (~70\%), yielding approximately 25--26 discrete SpO\textsubscript{2} target samples per subject. This produces broad coverage of three clinically relevant severity classes: Normal (SpO\textsubscript{2} 96--100\%), Moderate (92--95\%), and Severe (88--91\%)~\cite{lingamoorthy2026hypoxspike}.

Each participant wore both a custom shoulder-mounted sensor and a gold-standard reference pulse oximeter. The reference device provides the ground-truth SpO\textsubscript{2} labels, while the shoulder device records raw two-channel (Red, IR) PPG signals from the deltoid together with 6-axis IMU data capturing 3D acceleration and gyroscopic motion. These measurements support both respiration-rate estimation and motion-artifact detection. Unlike opportunistic sleep-apnea datasets, where severe hypoxemia events are infrequent and unevenly distributed, this controlled protocol yields a more balanced label distribution across the full SpO\textsubscript{2} range.

\subsection{Preprocessing and Denoising of Data}
The raw optical data are acquired using the MAX30101 integrated optical sensor, which uses red (660 nm) and infrared (880 nm) LEDs to illuminate the deltoid and a matched photodetector to capture the reflected light. The photodetector samples the signal at 400 Hz with 18-bit ADC precision. To suppress high-frequency noise, the sensor performs on-chip averaging over 16 consecutive samples before transmission to the microcontroller, reducing the effective sampling rate to 25 Hz. The resulting two-channel (Red, IR) PPG signal is stored in binary format in local flash memory.

Software preprocessing begins with temporal alignment and signal-quality filtering. Samples collected when either the shoulder device or the reference pulse oximeter was not worn are removed (indicated by the reduced correlation between the red and infrared PPG channels). The reference device’s SpO\textsubscript{2} readings are upsampled using forward-fill interpolation to align with the shoulder-sensor data, and any remaining temporal offset is corrected by cross-correlating pulse rates derived from the two signals. Because the shoulder is a highly dynamic sensing site, motion artifacts are identified using the vector magnitude of the 3-axis accelerometer signal, and high-motion samples are discarded before SpO\textsubscript{2} estimation. Low-quality segments are further removed using a signal quality index computed from the skewness of the PPG signal over an 8-second rolling window.

For SpO\textsubscript{2} estimation, both red and infrared PPG channels are filtered with a third-order Butterworth band-pass filter with cutoff frequencies of 0.5 Hz and 3.0 Hz to remove baseline drift and high-frequency noise. A Chebyshev Type II band-pass filter is similarly applied to 4-second signal epochs used for SpO\textsubscript{2} estimation. Finally, to address the inherent class imbalance in SpO\textsubscript{2} data, the dataset is downsampled so that each class matches the least frequent one.

The dataset is split with 90\% of subjects allocated for training and validation (80/20 split within the pool) and the remaining 10\% held for testing. 

\subsection{Hardware Performance Evaluation}
The trained SNN shown in Fig.~\ref{fig:scnn_architecture} is deployed on the neuromorphic hardware. Multiple fixed-point precision configurations were evaluated: 8-bit, 12-bit, and 16-bit, providing a finer-grained analysis of the accuracy--resource trade-off for hypoxia detection on the hardware.

Table~\ref{tab:spo2_accuracy} reports per-fold software accuracy and the corresponding hardware accuracy after quantization and deployment at three precision levels. The 16-bit design achieves an average hardware accuracy of 85.8\%, a 3.3\% drop from the software average of 89.1\%. Reducing precision to 12 bits yields 83.92\% average accuracy, only 1.88\% below the 16-bit result, making it an attractive accuracy-cost trade-off. In contrast, the 8-bit design falls to 73.54\%, indicating that the precision is insufficient to reliably represent the neuron state variables and synaptic weights required for SpO$_2$ classification. Fold~2 consistently achieves the best hardware accuracy across all precisions, matching its strongest software baseline.

\begin{table}[t]
    \centering
    \caption{Per-fold classification accuracy (\%) for hypoxia detection.}
    \label{tab:spo2_accuracy}
    \begin{tabular}{lcccc}
        \toprule
        & & \multicolumn{3}{c}{\textbf{HW Accuracy}} \\
        \cmidrule(lr){3-5}
        & \textbf{SW Test} & \textbf{16-bit} & \textbf{12-bit} & \textbf{8-bit} \\
        & & $Q_{1.8}/Q_{7.8}$ & $Q_{1.8}/Q_{3.8}$ & $Q_{1.4}/Q_{3.4}$ \\
        \midrule
        Fold 1 & 87.7 & 85.8   & 84.5   & 72.9 \\
        Fold 2 & 94.8 & 92.9   & 92.5   & 86 \\
        Fold 3 & 86.5 & 86.2   & 84.3   & 72.9 \\
        Fold 4 & 88.4 & 88.4   & 81.9   & 73.3 \\
        Fold 5 & 88.2 & 88     & 88     & 88 \\
        \midrule
        Avg    & 89.1 & 88.26  & 86.24 & 78.62 \\
        \bottomrule
    \end{tabular}
    \vspace{-6pt}
\end{table}

Table~\ref{tab:spo2_utilization} reports the utilization of FPGA resources along with dynamic power consumption and energy efficiency metrics. SpO\textsubscript{2} classification requires more resources than the MNIST and FMNIST benchmarks, reflecting the larger network (64 output channels in Conv~2, two FC layers) and the three input channels. At 16-bit, the design occupies 51.73\% of LUTs and 19.64\% of FFs. Reducing to 12-bit decreases LUT usage to 36.34\% and FF usage to 15.32\%, while the 8-bit configuration yields 37.53\% LUT and 10.99\% FF utilization. BRAM and DSP usage remain unchanged between 16-bit and 12-bit, as both precisions require dedicated DSP multipliers, while the 8-bit configuration eliminates DSP usage entirely.

\begin{table}[t!]
    \centering
    \caption{Post-implementation FPGA resource utilization, dynamic power, and energy efficiency on SpO\textsubscript{2} at 100KHz.}
    \label{tab:spo2_utilization}
    \resizebox{\ifdim\width>\textwidth \textwidth\else \width\fi}{!}{%
    \begin{tabular}{l*{6}{c}}
        \toprule
        & \multicolumn{2}{c}{\textbf{16-bit}} & \multicolumn{2}{c}{\textbf{12-bit}} & \multicolumn{2}{c}{\textbf{8-bit}} \\
        \cmidrule(lr){2-3} \cmidrule(lr){4-5} \cmidrule(lr){6-7}
        \textbf{Metric} & Used & \% & Used & \% & Used & \% \\
        \midrule
        LUT  & 674{,}353 & 51.73 & 473{,}724 & 36.34 & 489{,}218 & 37.53 \\
        FF   & 512{,}013 & 19.64 & 399{,}347 & 15.32 & 286{,}652 & 10.99 \\
        BRAM & 313.5     & 15.55 & 313.5     & 15.55 & 313.5     & 15.55 \\
        DSP  & 2{,}246   & 24.89 & 2{,}246   & 24.89 & ---       & ---   \\
        \midrule
        Dynamic (W)               & \multicolumn{2}{c}{1.455} & \multicolumn{2}{c}{1.284} & \multicolumn{2}{c}{0.771} \\[2pt]
        MOPS/W                    & \multicolumn{2}{c}{6.60}  & \multicolumn{2}{c}{7.48}  & \multicolumn{2}{c}{12.45} \\[2pt]
        Energy/inference ($\mu$J) & \multicolumn{2}{c}{727.5} & \multicolumn{2}{c}{642.0} & \multicolumn{2}{c}{385.5} \\
        \bottomrule
    \end{tabular}%
    }
    \vspace{-12pt}
\end{table}

When operating at 100~KHz, the 8-bit configuration consumes 0.771~W with an energy efficiency of 12.45~MOPS/W and 385.5~$\mu$J per inference. The 12-bit configuration increases dynamic power to 1.284~W , and the 16-bit to 1.455~W. These results demonstrate that reduced-precision quantization provides substantial power savings: the 8-bit SpO\textsubscript{2} configuration achieves a 1.89$\times$ reduction in dynamic power and energy per inference compared to 16-bit, highlighting the effectiveness of mixed-precision deployment for power-constrained biomedical edge applications.

Because biomarkers such as SpO\textsubscript{2} need only be monitored at low rates, for example once per second, the hardware can operate at a correspondingly low clock frequency. In our design, 16-bit operation at 10~kHz—--still far above what is required for 1~Hz monitoring—--consumes 148~mW of dynamic power, while reducing the frequency to 1~kHz lowers the dynamic power to 16~mW.

\subsection{Comparison with State-of-the-Art}
Table~\ref{tab:soa_comparison} compares the proposed framework with existing approaches along two axes: PPG-based hypoxemia classification and FPGA-based spiking neural network accelerators.

\begin{table}[t]
    \centering
    \caption{Comparison with state-of-the-art.}
    \label{tab:soa_comparison}
    \resizebox{\ifdim\width>\textwidth \textwidth\else \width\fi}{!}{%
    \begin{tabular}{lccccccc}
        \toprule
        \textbf{Work} & \textbf{Platf.} & \textbf{Model} & \textbf{CNN} & \textbf{Mixed} & \textbf{SpO\textsubscript{2}} & \textbf{Open} & \textbf{HW} \\
        & & & & \textbf{Prec.} & & \textbf{Src.} & \textbf{Depl.} \\
        \midrule
        OxyCaps~\cite{lingamoorthy2024drug}          & SW   & Capsule    & \checkmark & --- & \checkmark & \checkmark & $\times$   \\
        HypoxSpike~\cite{lingamoorthy2026hypoxspike}        & SW   & Tern. SNN  & \checkmark & --- & \checkmark & \checkmark & $\times$   \\
        Spiker~\cite{carpegna2022spiker}           & FPGA & SNN (FC)   & $\times$   & $\times$ & $\times$ & \checkmark & \checkmark \\
        DeepFire~\cite{aung2021deepfire}         & FPGA & SNN (CNN-FC)  & \checkmark & $\times$ & $\times$ & $\times$   & \checkmark \\
        DeepFire2~\cite{aung2023deepfire2}        & FPGA & SNN (CNN-FC)  & \checkmark & $\times$ & $\times$ & $\times$   & \checkmark \\
        NeuroCoreX~\cite{gautam2025neurocorex}       & FPGA & SNN (FC)   & $\times$   & $\times$ & $\times$ & \checkmark & \checkmark \\
        \midrule
        \textbf{Ours}    & \textbf{FPGA} & \textbf{SNN (CNN-FC)}    & \checkmark & \checkmark & \checkmark & \checkmark & \checkmark \\
        \bottomrule
    \end{tabular}%
    }
    \vspace{-12pt}
\end{table}
    
Among existing PPG-based hypoxia classifiers, OxyCaps~\cite{lingamoorthy2024drug} employs capsule networks and HypoxSpike~\cite{lingamoorthy2026hypoxspike} introduces a ternary spiking neural network, both of which achieve competitive classification accuracy in software. However, neither work has been deployed on neuromorphic hardware, and HypoxSpike explicitly identifies hardware deployment as a direction for future work. Among FPGA-based SNN accelerators, DeepFire~\cite{aung2021deepfire} and DeepFire2~\cite{aung2023deepfire2} support spiking CNN inference but lack mixed-precision quantization support and have not been applied to biomedical signal classification. Spiker~\cite{carpegna2022spiker} and NeuroCoreX~\cite{gautam2025neurocorex} are limited to fully connected topologies without convolutional layer support. To the best of our knowledge, the proposed framework is the first to combine a reconfigurable hybrid CNN--FC spiking architecture with variable-precision quantization and demonstrate end-to-end hardware-accelerated inference for SpO\textsubscript{2} classification from PPG signals.

\subsection{Limitations and Proposed Improvements}
While the proposed framework demonstrates the feasibility of hardware-accelerated SpO\textsubscript{2} classification, the reported dynamic power reflects the overhead of FPGA-based prototyping. A corresponding ASIC implementation, the natural next step for this architecture, is expected to substantially reduce power consumption while preserving the same configurability and accuracy. Additionally, all experiments used 50 simulation time steps per inference; reducing the number of time steps through adaptive temporal coding or training-aware optimization would proportionally lower the dynamic power and energy per inference. Future work will also explore deeper network architectures with pooling and on-chip learning capabilities, along with integration with wearable PPG sensing and event-driven neuromorphic sensors to enable end-to-end low-power SpO\textsubscript{2} monitoring at the edge.



\section{Conclusion}\label{sec:conclusion}
\label{sec:conclusion}
We present a configurable digital architecture for spiking convolutional neural network acceleration, targeting hypoxia detection as a representative edge application. The architecture extends a programmable, quantized, layer-based neuromorphic core with spiking convolutional support, enabling hybrid CNN--FC topologies with multi-level configurability. An snnTorch-based hardware--software co-design interface allows trained parameters to be deployed directly onto the hardware with full quantization support.

The design is validated on the MNIST and FMNIST benchmarks to show generalizability, achieving up to 98.0\% hardware accuracy on MNIST and 86.0\% on FMNIST at 16-bit precision. It is also evaluated on hypoxia detection formulated as a classification task using PPG signals and skin-tone information, achieving an average hardware accuracy of 85.8\% at 16-bit precision. Across all tasks, lower-precision quantization consistently improves energy efficiency, with 12-bit configurations reducing dynamic power by up to 1.13$\times$ relative to 16-bit implementations. To the best of our knowledge, this is the first demonstration of hardware-accelerated spiking CNN inference for hypoxia detection and classification.o support reproducibility, the complete HDL design
files, trained models are released as an open repository.\footnote{\url{https://github.com/Sarah-Johari/SCNN}}


\bibliographystyle{IEEEtran}
\bibliography{ref}
\end{document}